\documentclass[10pt,twocolumn,a4paper]{article} 
\usepackage{latex8}
\usepackage{times}
\usepackage{graphicx}

\begin{document}
\title{Extending the Trusted Path in Client-Server Interaction}

\author{Hanno Langweg and Tommy Kristiansen \\
Norwegian Information Security Laboratory -- NISlab \\
Department of Computer Science and Media Technology, Gj{\o}vik University College \\
P.O. Box 191, 2802 Gj{\o}vik, Norway \\
hanno.langweg@hig.no \\
}

\maketitle
\thispagestyle{empty}

\begin{abstract}
We present a method to secure the complete path between a server and the local human user at a network node. This is useful for scenarios like internet banking, electronic signatures, or online voting. Protection of input authenticity and output integrity and authenticity is accomplished by a combination of traditional and novel technologies, e.g., SSL, ActiveX, and DirectX. Our approach does not require administrative privileges to deploy and is hence suitable for consumer applications. Results are based on the implementation of a proof-of-concept application for the Windows platform.
\end{abstract}

\section{Introduction}
Interacting with the local human user is the weak point in client-server communications. While machines can employ cryptographical mechanisms to ensure authenticity, integrity, and confidentiality of communication, humans are not capable of this. They rely on their local computer to present data and transmit their input to a server reliably.

Today's operating systems provide protection against unauthorized modification of operating system components and offer mechanisms like discretionary access control and process separation to users and processes. Often, all processes of the same user operate with the same privileges. Malicious software (malware) can exploit this fact to read input destined for other processes (e.g. a keylogger) or modify the output displayed to the user (e.g. local phishing attack). Some banks in South Korea already -- including, e.g., Korea Exchange Bank and Woori Bank -- use ActiveX-based tools to prevent the successful use of keyloggers during internet banking, apparently after large-scale keylogging attacks in internet caf\'es some years ago.

A server application needs a trusted path to the user at a network node. This concept is not new and exists in operating systems. The secure attention sequence Ctrl+Alt+Del in Microsoft Windows is an example of how the user can invoke a trusted path to the operating system to log on. Output of a trusted path cannot be manipulated by other processes and input cannot be read. The process using a trusted path can be sure that input and output are shared only with the user. This authenticity is important when using all kinds of transaction systems, e.g. creating electronic signatures or online voting.

We present related work in the next section, followed by an analysis of the entities involved in secure client-server communication. We then outline secure output and secure input in sections \ref{section:output} and \ref{section:input}. Security of the implementation is discussed in detail in section \ref{section:security}. We conclude by examining mechanisms aimed to improve protection against malware in Windows Vista and by taking a look at other platforms.

\section{Previous and related work}
User interface security has always been an issue. Security evaluation criteria like the TCSEC \cite{TCSEC}, CTCPEC \cite{CTCPEC} or the Common Criteria \cite{CC2.1-2} require a Trusted path to establish a secure communication between the user and the operating system. The TCSEC defines it as follows: \emph{`Trusted Path -- A mechanism by which a person at a terminal can communicate directly with the Trusted Computing Base. This mechanism can only be activated by the person or the Trusted Computing Base and cannot be imitated by untrusted software.'} (p. 113)

A proposal for a user interface for \emph{SMITE} prevents Trojan horses from tampering with application output. \cite{Wiseman88} Kernelizing the graphics server and delegating window manager tasks to the application level is a prototypical solution in \cite{Feske05}. However, it is not compatible with the Windows platform used on the vast majority of existing client computers.

In the Microsoft Windows operating system, applications typically receive information about user actions by messages. Since these can be sent by malicious programs as well, they are a convenient attack vector. It is a vulnerability by design -- Windows treats all processes equally that run on the same desktop. If one needs an undisturbed interface, a separate \emph{desktop} attached to the interactive \emph{window station} should be assigned. That approach is pursued by \cite{Balfanz01}. However, managing separate desktops can be cumbersome for software developers. So most of today's software that interacts with a local user runs in a single desktop shared by benign and malign programs.

This problem is encountered by local security applications such as electronic signature software \cite{Spalka02b}, virus scanners, personal fire walls etc. In \cite{Schmid02} a dilemma is pointed out when notifying users about security events. Users are notified about presence of a possibly malicious program that could hide that very notification immediately. Some improvements to dialog-based security are shown in \cite{Carlisle01}. Application output should be defended against hiding. Actions should be delayed so that users could interfere when a program is controlled by simulated input or scripting. \emph{DirectX} can be used to achieve undisturbed output instead of the co-operative Windows GDI. \cite{USPatent6731756}, \cite{Langweg02}, \cite{Langweg04c} Modifying the web browser to convey meta-information to the user about which window can be trusted is advocated by \cite{Ye02}.

Window messages in event-driven systems in general are discussed in \cite{Xenitellis02a} and \cite{Xenitellis02b} where a lack of authentication is lamented. Rigorous filtering of messages is proposed. A straight-forward alternative, outlined by \cite{Spalka02c}, is to add an authenticated origin to messages. It requires changes in the decade-old backwards-compatible messaging system and is hence unlikely to be adopted by Microsoft. In the X-Windows system, a radical approach is pursued, allowing to disable transmission of messages by the \emph{SendEvents} function. There may be occasions, like computer-based training, in which remote control of another application or parts of it is desired. Only some applications expose an interface by which they can be automated explicitly (e.g., via the Microsoft Accessibility API). Consequently, simulating user input is a quick and convenient way for small helper applications. A reminder that messages can be sent between processes running in different security contexts provide \cite{Paget02} and \cite{Howard02}: \emph{`In the Windows user interface, the desktop is the security boundary, and any application running on the interactive desktop can interact with any window on the interactive desktop, even if that window is invisible. This is true regardless of the security context of the application that creates the window and the security context of the application.'} DirectX can ameliorate problems with forged messages. \cite{Langweg04c} We found one company reporting to use SSL to encrypt input between the input device and the application. They claimed they were using patented technology from Korea, but were unwilling or unable to provide any patent number or technical information on their product. \cite{LocalSSL06}

Digital rights management (DRM) techniques also intend to preserve integrity and confidentiality of displayed information. However, the threat scenario is different. DRM has to protect content also against the local user, we need protection of the user against malware actions.

As is often the case, the problem could be solved by using a separate hardware device to display data and receive user input. It would be just a physical representation of restricted access to resources and processes. Unfortunately, hardware incurs additional costs and does not lend itself to fast and massive roll-out to a large user base.

\section{Client-server interaction}
\begin{figure}[h]
   \centering
   \includegraphics[width=0.45\textwidth]{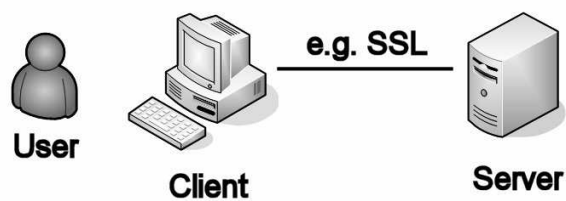}
   \caption{Client-server interaction.}
   \label{figure:clientserver}
\end{figure}

A number of applications today are structured after the client-server pattern: internet banking, contract signing, e.g. in e-government, or online voting. Here, the main application is run on highly protected servers. Users connect to the server from their local machine. The machine acts as a smart terminal, collecting user input, transmitting it to the server, receiving server data and displaying server output. The local user initiates and completes transactions with the server application. This is depicted in figure \ref{figure:clientserver}.

The server application does not interface directly with the user. It connects to a client application; probably it is an application (e.g. a web page, Java applet, or ActiveX control) that was sent by the server in advance. This client application runs alongside other processes. Most of these are benign processes which do not interfere with the client-server interaction, while some might have a malicious purpose.

\begin{figure}[h]
   \centering
   \includegraphics[width=0.45\textwidth]{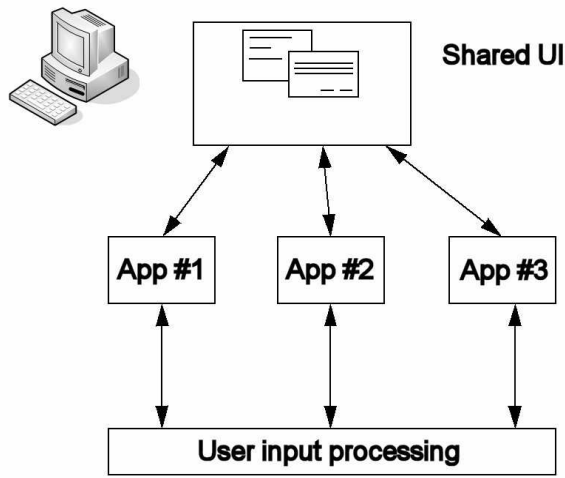}
   \caption{Local user interface architecture.}
   \label{figure:localapps}
\end{figure}

Processes are separated from each other by the client operating system. They share resources, most notably files, memory, and the user interface. Often, they are executed in the same security context so that access control cannot be used to distinguish different privileges. See figure \ref{figure:localapps}.

The user interacts with a local application via the local user interface. Some problems immediately arise:
\begin{enumerate}
   \item How do user and application know which server they are talking to?
   \item How does the server know which application it is talking to?
   \item How does the user know which application input is directed to?
   \item How does the user know which application produces the output?
   \item How does the application know that user received the output?   
   \item How does the application know where input comes from?
\end{enumerate}

The first two problems can be solved by using a cryptographic protocol that offers secure authentication of the communicating parties and integrity of the communication, e.g. SSL. The strength of the cryptographic algorithm relies on access of the adversary to encrypted data and on it being computationally infeasible to decrypt the data or forge a digital signature.

The remaining four questions demand a \emph{trusted path} between the local application and the user. The local user interface is the weak link in the interaction of the user with the server application. An adversary is much more likely to attack here than spending resources on breaking a cryptographic algorithm -- breaking cryptography is typically either a formidable mathematical challenge or requires a large amount of computing resources. Attacks on the server are another option. However, a server is usually easier to protect than a large number of clients. Server protection is outside the scope of this article.

Malware may act completely autonomous or may have a communication channel to a human agent. The human agent could remotely control the malicious process or assist it in analysing the situation in real time. An adversary could launch a man-in-the-middle attack, trying to open a session with the server and then mimicking the server towards the client application. It could also try to simulate user input, capture user input used for authentication, manipulate the output to mislead the user or cover up input manipulation.

In our analysis we focus on Microsoft Windows~XP/2000 and present an implementation of our local trusted path. Windows~XP is the system most likely in use at most clients. We cover the successor of Windows~XP -- Vista -- and also look at alternative platforms.

We assume that the user works as a standard user and is not running malware processes under an administrative account. In that case we would not have to worry about a trusted path, since the malicious process would find more attractive targets exploiting its administrative rights.

\section{Secure client output}
\label{section:output}
Security of output to a window on a shared Microsoft Windows desktop is limited. There is no confidentiality; all processes attached to the desktop can capture the desktop's content and that of all windows on the desktop. There is no integrity; all processes can modify the desktop's content and that of all windows. \cite{Yuan06} It is a limitation by design as all processes sharing a desktop are assumed to behave cooperatively. If no cooperation can be vouched for, then separate desktops are generally recommended to solve this problem. The limitation described applies only when the GDI (Graphics Device Interface) is used.

Another option is to use Microsoft \emph{DirectX} as explained in \cite{Langweg02} and \cite{Langweg04c}. Microsoft DirectX is a group of technologies designed by Microsoft for running games. It is an integral part of Windows~XP (since Windows~98). DirectX gives software developers a consistent set of APIs that give improved access to hardware. These APIs control low-level functions, including graphics memory management and support for input devices. Of the various components, DirectDraw/Direct3D is responsible for output devices, DirectInput addresses input devices. DirectDraw allows to access the display hardware in exclusive full screen mode, keeping other programs from distorting the information presented to the user.

\begin{figure}[h]
   \centering
   \includegraphics[width=0.45\textwidth]{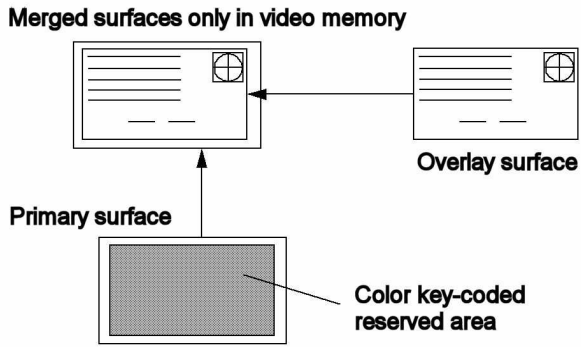}
   \caption{Combining overlay surface and primary surface.}
   \label{figure:overlay}
\end{figure}

DirectDraw works with surfaces on which processes can draw. One such surface is the GDI surface that is shared with the other processes in the current desktop. Surfaces can exist in system or video memory. For increased security, video memory should be used. Screen grabber software can read from system memory, but not from video memory. To avoid transfer of the content to system memory when there is not enough video memory available, we use an overlay surface. Overlay surfaces are merged with the primary surface representing the screen's content. The overlay surface's content is shown at all places on the primary surface where the color key is used -- one of the colors in the surface's palette of available colors. Pixels in the color key's color are replaced with pixels from the overlay surface, as shown in figure \ref{figure:overlay}. This does not modify the primary surface in any way. A screen grabber program that might gain read access to surfaces in system memory can only capture the primary surface's content. The user meanwhile sees the content of the overlay surface that only exists in video memory. Combination of the surfaces is done internally by the video graphics adapter and, hence, cannot be tampered with. We achieve integrity and confidentiality of screen output.

Now that we have integrity, we need our control to authenticate to the user. In principle every process is capable of requesting fullscreen exclusive access to the screen. We need to show the user a secret that only the server-provided process and the user share. This technique is known as \emph{window personalisation}.

As a shared secret we use what we call an \emph{application hologram}. It is basically a sequence of related images shown in an animation. It could also be a single image, but an animation is more likely to get the user's attention. The user is also more likely to notice its absence. The hologram is made known to the user upon registering with the service. Here, the same process can be used that also distributes a PIN code, a list of transaction numbers, or a security token to the user. The hologram could also be selected or uploaded by the user.

The hologram is transferred by the server to the ActiveX control. For this purpose, an SSL connection is used. The control uses a client key provided by the server when the control was downloaded in the web browser. In this way the server can verify that the hologram is requested by the recently downloaded control instead of malware running on the same host. The control is protected against accesses from other processes by the process separation techniques of the operating system. Hence, the client key and the hologram are only known to the ActiveX control.

Hence, we can show arbitrary data in exclusive fullscreen mode to the user. It cannot be modified or captured by other processes. The user can determine the data's origin by verifying that the correct application hologram is present.

\section{Secure client input}
\label{section:input}
We look at authenticity/integrity and confidentiality of user input. Input is affected by window messages, DirectInput, SendInput, and low-level hooks.

Microsoft Windows uses an internal messaging model to control Windows applications. Messages are generated whenever an event, e.g. a key press or a mouse move, occurs. However, not just user actions induce messages. Malicious software can construct and send messages, thereby simulating user input to dialog components. It is not possible for a program to distinguish between messages placed in the queue by the operating system and messages placed by another application. \emph{DirectInput} retrieves information before it is distilled by the operating system to Windows messages. Hence, input synthesized by placing a forged message in a program's message queue can be detected and ignored. The \emph{GetAsyncKeyState} function provides a similar way to verify if a key has been pressed/released or if only a message has been sent.

If messages are not forged, input can be simulated by the \emph{SendInput} function. Key presses/releases and mouse movements can be placed in the raw input queue. This queue is used by both GetAsyncKeyState and DirectInput. Input simulated this way appears in the same way as genuine input.

Low-level keyboard hooks have access to the raw input queue. Here, simulated input is marked with an "injected" flag. An application using this type of hook can thus detect and discard simulated input. There are two drawbacks: low-level keyboard hooks can be used to capture input to other processes, e.g. by a keylogger, and other processes can install these hooks and remove the flag. A method to counter this is to renew the hook permanently. The newest hook is placed first in the hook chain and gets to process the input data before it can be tampered with.

It may be possible to distinguish users and untrustworthy programs by observing their input behaviour, e.g. programs simulating input much faster than an ordinary user could type. This rather falls in the field of behavioral biometrics. \cite{Bergadano02}

Our solution uses \emph{DirectInput} or the Win32 API function \emph{GetAsyncKeyState} (two implementation variants) to cope with forged window messages. Defeating SendInput and low-level keyboard hooks requires fast renewal of our own low-level keyboard hook, e.g., every 100~ms.

Hooks are troublesome for another reason. A system-wide hook is called by all processes in the same desktop. An attacker could use a hook not only to process events, but also to execute code of their choosing in the address space of another process. Hooks can be disabled for a desktop session, but then we also lack the possibility to detect simulated input.

We found a solution for this in combination with undisturbed output. As \cite{Gobioff96} shows, trusted output plus a single bit of trusted input is equivalent to trusted input. User input -- genuine or simulated -- is collected by our ActiveX control and sent to the server. The server stores the data in preparation of the next transaction and echoes it back to our control, adding a unique random value. The control shows the data on its secure surface, together with its application hologram and the random value. It then asks the user for confirmation. The user checks whether the data for the transaction are correct and if the application hologram is present. In the positive case, the user inputs a one time password based on the random value. In the negative case, the user closes the session or otherwise aborts communication with the control and the server -- a single bit of trusted input.

The one time password based on the random value can be implemented in various ways. We use a simple list with transaction numbers (similar to those used by some banks for their online banking) where the user looks up the transaction number associated with the random value. It could also be computed by a token that shows a confirmation number upon entering a challenge. The token then basically stores or computes a list with transaction numbers and outputs only the number associated with the given random value.

Hence, we cannot prevent malware from simulating user input, but we can prevent simulated user input from having an effect on the transaction.

\section{Client security requirements}
\label{section:security}
\paragraph{Operating system integrity --}
The security of our implementation of a trusted path for an application relies on the integrity of the operating system.

We have analysed which operating system modules are needed to download and execute our ActiveX control, and which are involved in using DirectX. The operating system is responsible for the integrity of its components. Windows employs discretionary access control (DAC) and integrity monitoring and restore of system files (WFP). Use of file protection mechanisms is shown in table \ref{table:fileprotection}. Most files reside in the \emph{Windows} or \emph{Program Files} folder (or subfolders). All are protected by access control entries preventing modification by processes executing under a standard user account. Some files enjoy protection by the Windows file protection service in addition. None of the files used by Internet Explorer or DirectX is unprotected against malware in a standard user account.

\begin{table}
\begin{center}
\begin{tabular}{lrrrr}
\emph{Usage}      & \emph{Unprotected} & \emph{DAC} & \emph{WFP} & Total \\
\hline
Internet Explorer &                 0  &        112 &         24 & 112 \\
DirectX           &                 0  &        104 &         20 & 104 \\
\hline
\end{tabular}
\caption{\label{table:fileprotection} Used files and their protection mechanisms.}
\end{center}
\end{table}

\paragraph{Non-developer system --}
DirectX comes in a run-time version used on most systems and in a developer version. While in the run-time version exclusive access provides confidentiality of output, the developer version opens for screen capture by malicious software. Hence, we recommend installing the non-developer version of DirectX when using our solution. Integrity is preserved in both variants. Availability of the screen for output is handled on a first-come first-served basis. Security-sensitive applications are advised to request access in full screen mode early and use application holograms to prove their authenticity to the user.

\paragraph{Screen capture software --}
Programs that are able to capture DirectDraw surfaces do so by API hooking from an injected DLL. Using reverse engineering, we found that a DLL was injected in running processes to capture calls to the DirectDraw API. Once the adversarial code gets hold of the COM interface pointer, it can access the surfaces from the injected DLL and transfer screen contents to the capture software. Similar methods could be used by malware. It is imperative that other processes be prevented from injecting code. Hooking can be prevented by several means. The straight-forward approach is to disable systemwide hooks for the desktop object. This could be achieved during login. Another approach is to employ API hooking to restrict calls to the {\tt SetWindowsHookEx} function. Both techniques cannot be used from a server-provided ActiveX control. However, hooks need to be prevented anyway. A malicious process being able to execute code in the address space of another process can affect the system's security performance adversely. There are more attractive targets than performing (non-trivial) API hooking and screen capture. Analysing and modifying a surface reliably and undetected in real time poses a high bar for an attacker.

\paragraph{Man-in-the-middle-attack (MITM) --}
MITM attacks involving a spoofed server can be detected by use of the SSL protocol. The user can check the certificate of the server connected to.

\paragraph{User interface remote control --}
With the user interface not being visible to malware, the attacker has to guess how to simulate user actions. By observing mouse movements or key strokes it might be possible for the attacker to learn about the user interface straucture. We could vary the interface slightly from time to time. However, the attacker might as well simulate---the server collects the (manipulated) data, sends it back and displays the data via the trusted output, then asks for confirmation. If the data was manipulated, the user can cancel the transaction. If the information is correct, the user enters a one-time usable transaction number. The number is bound to the data received. We could also request the user to solve a CAPTCHA (Completely Automatic Public Turing test to tell Computers and Humans Apart) before entering confirmation. \cite{Ahn03} Malware would not be able to solve it and hence, premature confirmation of faked data could be detected and rejected.

Our solution hence bears the following security properties:
\begin{enumerate}
   \item We can prevent user interface modification (DirectDraw)
   \item We can detect user interface spoofing when it happens (Application hologram, SSL certificate)
   \item We can protect user interface confidentiality (overlay surface)
   \item We can detect user interface remote control after it has happened (window messages, transaction number list/token)
   \item We \emph{cannot} prevent malware from denying service
\end{enumerate}

Our focus is on a sound design of how protection mechanisms are used. If there are flaws owing to errors in the implementation of the platform, they have to be covered by other means. If one sees the Windows platform as an insecure platform because of implementation flaws, one should refrain from using any security-relevant software on it. However, with the current market share, we have to offer practical solutions that help users who do not want to install a different operating system.

\section{Outlook: Windows Vista}
\label{section:vista}
Vista -- the new Microsoft Windows operating system version -- proposes some interesting techniques to improve malware resistance. It is not yet clear if all the functionality currently present in the beta version will be in the final product. Documentation is still incomplete at best; it is mostly provided as short-lived weblog communication. For our purposes of achieving an unmanipulated user interface, \emph{MIC} (Mandatory Integrity Control), \emph{UIPI} (User Interface Privilege Isolation), \emph{UAC} (User Account Control), Internet Explorer 7 Protected Mode, and the \emph{Desktop Window Manager} are relevant.

With \emph{Mandatory Integrity Control} every process and every securable object are assigned an integrity level: \emph{low}, \emph{medium}, \emph{high}, \emph{system}. The level of the process must dominate the level of the object to be able to modify the object. MIC is applied before DAC (discretionary access control) permissions. It especially protects operating system files from modification. MIC in combination with \emph{User Interface Privilege Isolation} this prevents processes at a lower level to send window messages to processes at a higher level. In addition, it prevents hooks created by lower level processes to be called by higher level processes.

Processes started by the shell usually run at \emph{medium} integrity level. If our ActiveX control could be elevated to run at e.g. \emph{high} level, it would be protected from forged windows messages and message hooks. However, \emph{high} level might require running the control with administrative privileges, something we want to avoid.

\emph{User Account Control} helps to run processes with few privileges. Processes will execute with a restricted standard user access token. If an action requires more privileges, the operating system asks the user to provide proper credentials to complete the task. Malware is hence restricted in its action under all accounts. The operating system switches to the logon desktop to show the elevation prompt using the trusted path to the user. However, malware could pretend to show an elevation prompt since the user does not use a secure attention sequence and the prompt cannot prove its origin.

Internet Explorer 7 will run in so-called \emph{Protected Mode}. It is basically a process running at \emph{low} integrity level under a user account of its own. It has write access only to a small number of folders and registry locations. If Internet Explorer gets compromised, malware cannot spread to other locations in the system and is executed only under the restricted account.

The output model is going to change. Windows on the desktop are managed by the \emph{Desktop Window Manager} (DWM). When we acquire a device context for our DirectDraw surface to draw on, DirectDraw locks the surface. This causes the DWM to enter a compatibility mode and disable some of the fancy effects of the new user interface. The system is still functional, though.

The biggest improvement as regards a trusted path for applications will be \emph{User Interface Privilege Isolation}, limiting the use of system-wide hooks.

\section{Other operating platforms}
\paragraph{Java} offers fullscreen exclusive mode access to the screen starting with version 1.4. \cite{Java1.4} This functionality is procided by the class {\tt java.awt.graphicDevice}. Methods are very similar to the DirectDraw API. In fact, Java uses DirectDraw if it is available on the platform. Otherwise, fullscreen mode is simulated with a top-level window filling the whole GDI surface. On Unix platforms, Java may use \emph{X11}. No API is provided to check for the authenticity of input. It is not possible to use overlay surfaces that always are stored in video memory. Conventional surfaces used for drawing can be placed in system memory by the video driver.

The Java Native Interface allows access to platform-dependent code. In principle, a Java application could also use DirectDraw and DirectInput via this interface. If Java was used as a replacement for \emph{ActiveX}, the client security requirements would have to include securing the \emph{Java Virtual Machine} against manipulation by untrustworthy processes.

\paragraph{Qt} is a cross-platform development framework for Windows, Mac, and X11. \cite{Qt} If consists of more than 400 classes, many of which can be used for user interface construction. \emph{Qt} can also be used in conjunction with ActiveX. However, Qt renders to the primary GDI surface under Windows and leaves its output vulnerable to modification by other processes. It might be possible to subclass or wrap the Qt widgets to use DirectX instead. The single advantage we see with Qt widgets is their independence of window messages. Malware cannot simply modify or influence the user interface by sending window messages.

\paragraph{OpenGL} is an industry standard API for 2D and 3D graphics programming. \cite{OpenGL} It is used to render graphical objects to a frame buffer. Access to the buffer is controlled by the cooperative window manager. OpenGL can hence not guarantee exclusive access.

\paragraph{GTK+} is a multi-platform toolkit for creating graphical user interfaces. \cite{GTK} Like \emph{Qt} or \emph{OpenGL} it executes on top of the operating system and cannot assure exclusive access to user interface hardware.

\paragraph{Simple DirectMedia Layer (SDL)} is a cross-platform multimedia library designed to provide low level access to input and output hardware. \cite{SDL} Multiple platforms are supported, including Windows, Linux, and Mac. \emph{SDL} does not establish direct access by its own means, but leverages the capabilities of the operating system. Under Windows, DirectX is used if driver support is available.
There exist implementations that use \emph{SDL} with \emph{DirectFB} on the Linux platform.

\paragraph{Linux} offers neither \emph{ActiveX} nor \emph{DirectX}. Access to the screen is possible using the {\tt fbdev} frame buffer device. It is hardware-independent and provides screen access from the console. The frame buffer device is used by the API layers introduced by \emph{GTK+} or \emph{SDL}. The \emph{DirectFB} project \cite{DirectFB} seems to offer more control over frame buffer access. However, it has not yet reached a 1.0 version number, and it remains unclear what has to be installed before a server-provided local application can use the API.
If applications run under the X-Windows server, the X-Window windowing system is used. It manages cooperatively how windows are drawn on the screen. It is possible to restrict the use of the \emph{SendEvents} function to simulate user input.

\section{Conclusions}
Communicating with the local human user is the weak link in client-server interaction. We have presented a method to ensure the integrity and authenticity of client-server communication from end to end.

Our approach makes use of existing technology, i.e. SSL and ActiveX to deliver an application to the user's machine, and DirectX to securely present information. No modification of the operating system is needed and no expensive hardware is employed. Installation can be done on-the-fly and, hence, can be centrally managed and supported. Instant delivery also ensures that the process can bear a server-implanted secret to authenticate to the server.

We provide components that application developers can use in their own projects. They can be used to retrofit existing ActiveX controls with exclusive screen access and process authentication. They can also be an example for developers including this technology in their applications from the ground up.

Further research should include how the new version of Microsoft Windows impacts the techniques and how other operating systems could be upgraded to provide secure local interaction. How to prevent denial-of-user-interface-access by local malware is also an open question.

\end{document}